\documentclass{epl}

\title{Magnetic energy of a quantum current}

\author{F. Miglietta \thanks{E-mail: \email{francesco.miglietta@pv.infn.it}}}

\institute{Dipartimento di Fisica Nucleare e Teorica dell' Universit\`a
di Pavia - via Bassi 6, \\ I - 27100 Pavia - Italy 
\\ I.N.F.N. - Sezione di Pavia}
\begin{document}

\maketitle

\begin{abstract}
It is shown that the magnetic energy of a quantum current, contrary to the
classical case, is essentially negative. Since this result allows to escape a
famous theorem by Bloch, it can be expected that, under appropriate conditions,
the ground state of a quantum conductor may be characterized by a spontaneous
current. A similar idea was suggested, many years ago, by Heisenberg and by
Born and Cheng as a possible mechanism for superconductivity.
\end{abstract}

\section{Introduction}

It is well-known, from the study of the infra-red divergence problem in
relativistic QED, that a charged particle cannot be separated from its {\it
classical} field \cite{1} \cite{2}. It is known also that the motion of a
classical particle cannot be described correctly, if the interaction with the
self-generated field is not taken into account properly \cite{3}.  \par 

In this paper we analyse the effects induced by the self-generated {\it
classical} ({\it i.e.} coherent) field on a stationary current flowing in a
macroscopic  quantum conductor. In particular we will estimate   the
contribution to the total energy of the system due to the {\it classical}
magnetic field and to its interaction with the current. The analysis will
concern explicitly two geometrically different models. In both cases we will
obtain the remarkable result that the magnetic energy associated to the current
is essentially negative. A completely different result would be obtained for a
current due to a system of classical charged particles. Such a difference
descends from the fact that, in classical mechanics, the kinetic energy of a
particle is expressed, in a natural way, in terms of the velocity. We remark
that the velocity is essentially a classical concept. In quantum mechanics the
velocity of an electron corresponds to the following operator

\begin{equation}
\bm{v} = m^{-1} [- i \hbar \bm{ \nabla } + e {\bm{A} }_c ( t,\bm{r} )]
,
\end{equation}
where $ {\bm{A} }_c $ is the {\it classical} vector potential. Owing to the
presence of $ {\bm{A} }_c \, $, which depends on the other particles of the
system as well as on the external environment, the velocity $ \bm{v} $, given
by eq.(1), is essentially a collective operator. On the other hand, the
canonical momentum is connected directly to the particle wave-length, which in
turn is involved by a boundary condition like periodicity. For these reasons it
looks more proper, in our opinion, to analyse a quantum many-body system in
terms of the canonical momenta and not in terms of the velocities of the
particles. \par

As previously mentioned, in this paper we will obtain the result that the
magnetic self-interaction energy of a quantum current is essentially negative.
This result may have unexpected consequences concerning {\it e.g.} the ground
state of a system of electrons subject to periodic boundary conditions. In
fact, as it will be discussed in the sequel, it is conceivable that, under
appropriate conditions, a physical situation corresponding to a spontaneous
current may be favoured with respect to a current-less situation. We recall
that attempts to explain superconductivity in terms of spontaneous currents
have been made long ago by Heisenberg \cite{4} and by Born and Cheng \cite{5}.
However such theories have been ruled out by a theorem by Bloch \cite{6}, owing
to the lack, in our opinion, of a proper analysis of the magnetic
self-interaction.   \par

The Schr\"odinger equation for a quantum system interacting with its {\it
classical} field is given by

\begin{equation}
i \hbar {\partial }_t \vert \psi \rangle = H[{\bm{A} }_c ] \;  \vert \psi
\rangle ,
\end{equation}
where 

\begin{equation}
H[{\bm{A} }_c ] = H_M - \int d \bm{r} \, {\bm{A} }_c \cdot ({\bm{j} }_p
+ \frac{1}{2} \: {\bm{j} }_A )  .
\end{equation}
We have assumed the Coulomb gauge. In eq.(3) $ H_M $ represents the
Hamiltonian for the matter, including the Coulomb interaction among charged
particles. In the interaction term the current-density operator $ \bm{j} $
has been splitted into two contributions. The first one, which will be called
{\it canonical} current, consists of

\begin{equation}
{\bm{j} }_p ( \bm{r} ) = - i \hbar {\sum }_a (q_a / 2 m_a) [\delta (
\bm{r} - {\bm{r} }_a ) {\bm{\nabla } }_a + {\bm{\nabla } }_a\delta (
\bm{r} - {\bm{r} }_a ) ] .
\end{equation}
The second contribution, involving the potential $ {\bm{A} }_c $,
consists of

\begin{equation}
{\bm{j} }_A ( \bm{r} ) = - {\sum }_a ( q^2_a / m_a ) \: n_a (\bm{r} ) \,
{\bm{A} }_c (\bm{r} ) ,
\end{equation}
where $ n_a (\bm{r} ) = \delta ( \bm{r} - {\bm{r} }_a )$ represents the
density operator for the a-th particle. A third contribution due to the spin
magnetization-density has been neglected, for the sake of simplicity. \par
   The {\it classical }potential $ {\bm{A} }_c $ is solution to the Maxwell
equation

\begin{equation}
\bm{\nabla } \times {\bm{B} }_c -c^{-2} {\partial }_t {\bm{E} }_{\perp c}
= {\epsilon }_0^{-1} c^{-2} \langle \psi \vert ({\bm{j} }_p + {\bm{j} }_A )
\vert \psi {\rangle }_{\perp } ,
\end{equation}
with the transverse electric field  given by

\begin{equation}
{\bm{E} }_{\perp c} = - {\partial }_t {\bm{A} }_c .
\end{equation}
Both eq.(6) and eq.(7) can be obtained, in the framework of QED, by use of the
coherent-state formalism \cite{7}. \par
   
In this paper we limit ourselves to consider the interaction of the system with
the {\it classical} field, neglecting any residual interaction with the quantum
field (uncoherent photon emission and reabsorption). In this approximation, a
simple expression for the conserved total energy of the whole system (matter
plus {\it classical} field) can be given, in the form:

\begin{equation}
{\cal E } = \langle \psi \vert H[ {\bm{A} }_c ] \vert \psi \rangle +
({\epsilon }_0 / 2 ) \int d \bm{r} \, ( \vert  {\bm{E} }_{\perp c} {\vert
}^2 + c^2 \vert {\bm{B} }_c {\vert }^2 ).
\end{equation}

It can be verified easily that $ {\cal E} $ is conserved. In fact, by
differentiating both sides of eq.(8) with respect to time, we obtain, in the
absence of emission of radiation,

\begin{eqnarray} 
&& \frac{d{\cal E} }{dt} = - \frac{i}{\hbar } \langle \psi \vert [ H[ {\bm{A}
}_c ] , H[ {\bm{A} }_c ] ] \vert \psi \rangle   + \int d \bm{r} \,
\frac{\partial {\bm{A} }_c}{\partial t} \cdot \{ - \langle \psi \vert ( {\bm{j}
}_p + {\bm{j} }_A ) \vert \psi \rangle \\ \nonumber 
&& - {\epsilon }_0 \frac{\partial {\bm{E} }_c }{\partial t} + {\epsilon }_0
c^2 \bm{\nabla } \times {\bm{B} }_c - {\epsilon }_0 c^2 \bm{ \nabla }
\cdot ( {\bm{E } }_c \times {\bm{B} }_c ) \}    
 = - {\epsilon }_0 c^2 \oint d \bm{s} \cdot ( {\bm{E} }_{\perp c } \times
{\bm{B} }_c )   = 0 ,
\end{eqnarray}
where eq.s (2) and (6) have been used. \par 

The second term in the r.h.s. of eq.(8) represents the (positive) energy of the
{\it classical} field. By a simple calculation, this term can be cast in the
form

\begin{eqnarray}
&& \frac{{\epsilon }_0}{2} \int d \bm{r} \, ( \vert {\bm{E} }_{\perp c}
{\vert }^2 + c^2 \vert {\bm{B} }_c {\vert }^2 )  
 = \frac{1}{2} \int d \bm{r} \, {\bm{A} }_c \cdot \langle \psi \vert ({\bm{j}
}_p + {\bm{j} }_A ) \vert \psi \rangle  + {\epsilon }_0 \int d \bm{r} \, \vert
{\bm{E} }_{\perp c} {\vert }^2    \\ \nonumber
&& -\frac{{\epsilon }_0}{4} \frac {d^2}{dt^2}
\int d \bm{r} \, \vert {\bm{A} }_c {\vert }^2 + \frac{1}{2} {\epsilon }_0 c^2
\oint d \bm{s}  \cdot ( {\bm{A} }_c \times {\bm{B} }_c )  ,
\end{eqnarray}
where eq.s (6) and (7) have been used. In the absence of a significant emission
of  coherent radiation the last term can be neglected, for a finite system. By
use of this result we obtain from eq.(8) 

\begin{equation}
{\cal E} = \langle H_M \rangle - \frac{1}{2} \int d \bm{r} \, {\bm{A} }_c \cdot
\langle {\bm{j} }_p \rangle  + {\epsilon }_0 \int d \bm{r} \, \vert { \bm{E}
}_{\perp c} {\vert }^2 - \frac{{\epsilon }_0}{4} \frac{d^2}{dt^2} \int d \bm{r}
\, \vert {\bm{A} }_c {\vert }^2 .
\end{equation}
We notice the absence of any contribution due to $ {\bm{j} }_A $ . We notice
also the factor of $ 1/2 $, as well as the minus sign, in the second term in
the r.h.s. of eq.(11). In stationary conditions, eq.(11) reads

\begin{equation}
{\cal E} ={\cal E}_M +{\cal E}_F = \langle H_M \rangle - \frac{1}{2} \int d
\bm{r} \, {\bm{A} }_c \cdot \langle {\bm{j} }_p \rangle ,
\end{equation}
where the Coulomb gauge is understood. The analysis of the physical content of
eq.(12), in particular of the minus sign in the r.h.s., can be simplified if
the potential $ {\bm{A} }_c $ is expressed in terms of the {\it canonical}
current $ {\bm{j} }_p $ and not in terms of the total current appearing in the
r.h.s. of the Maxwell equation (6). This result will be obtained explicitly for
two models.  \par

\section{First model}

The first model consists of a hollow cylinder with a current flowing around it.
Let $ L $ be the height, $ R $ the inner radius, $ a $ the thickness and let us
assume, in order to simplify the calculations, $ a \ll R \ll L $. Let us assume
an uniform {\it canonical} current, flowing around the cylinder, given by

\begin{equation}
\langle {\bm{j} }_p \rangle = -en ( \hbar p / m )
= -en ( \hbar \mu / mR ) ,
\end{equation}
where $ n $ is the (constant) electron density. We observe that $ \mu $ is
related to the total orbital angular momentum $ \hbar M $ of the electrons by
the relation $ \mu = MN^{-1} $, where $ N = 2 \pi RaLn $ represents the total
number of electrons involved in the process. \par
   
From eq.(6) we obtain the following expression for the magnetic flux $ {\Phi
}_B $

\begin{equation}
{\Phi }_B \simeq - h e^{-1} \mu [ 1 + (2/{\gamma }^2 aR ) ]^{-1}  \simeq 
- he^{-1} \mu,
\end{equation}
where $ \gamma $ is given by

\begin{equation}
\gamma = \sqrt{e^2 n / {\epsilon }_0 m c^2 }  . 
\end{equation}
For copper {\it e.g.} $ {\gamma }^{-1} \simeq 1.8 \, \times \, 10^{-8} $ m (in
the theory of superconductivity $ {\gamma }^{-1} $  is known as {\it
penetration depth} \cite{8}). The last expression in eq.(14), corresponding to
the case in which $ {\gamma }^2 aR \gg 1 $, is independent of geometrical
details. The following relation 

\begin{equation}
\langle j_A \rangle \simeq - ( e^2 n / 2 \pi R m ) {\Phi }_B
\end{equation}
has been used in the derivation of eq.(14). Eq.(14) gives an expression for
the flux $ {\Phi }_B $ in terms of $ \mu $, {\it i.e.} in terms of the total
angular momentum $ M = \mu N $.  \par

For this model the magnetic energy $ {\cal E}_F $, defined in eq.(12), is
negative:

\begin{equation}
{\cal E}_F \simeq - (1/2) \langle j_p \rangle {\Phi }_B \, aL  
= - N \frac{ {\hbar }^2 {\mu }^2}{2m R^2} (1 + \frac{2}{{\gamma }^2 aR }
)^{-1} \simeq - N \frac{ {\hbar }^2 {\mu }^2 }{2 m R^2 } \simeq  - \frac{N
{\hbar }^2 p^2 }{2 m} .
\end{equation}

\section{Second model}

Now let us consider a second model, consisting of a long, thin cylindrical
conductor, with a current flowing along it. Let $ L $ be the length and $ {\rho
}_0 $ the radius, with $ {\rho }_0 \ll L $. We assume an uniform stationary
{\it canonical} current, flowing inside the conductor, given by

\begin{equation}
\langle j_p \rangle = -e m^{-1} n \hbar p ,
\end{equation}
where $ \hbar p = N^{-1} \hbar P $ is the canonical momentum per electron (in
this case the total number of electrons is given by $ N = \pi {\rho }_0^2 L n
$). \par
   
For a very long conductor we may assume that $ {\bm{A} }_c $ is
approximately  oriented in the direction of the current (z-direction) and that
it depends, approximately, on the distance $ \rho $ from the axis of the
cylinder only. Putting $ x = \gamma \rho $, we obtain from eq.(6)   

\begin{equation}
{\nabla }^2 A_c - \theta ( x_0 - x ) A_c = \theta ( x_0 - x ) \hbar e^{-1} p .
\end{equation}
The solution to eq.(19) has the form

\begin{equation}
A_c = - \hbar e^{-1} p + \theta ( x_0 - x ) A_{int} + \theta ( x - x_0 )
A_{ext} ,
\end{equation}
where $ A_{int} $ is solution to

\begin{equation}
\frac{d^2 A_{int} }{d x^2 } + \frac{1}{x} \frac{d A_{int} }{d x} - A_{int} = 0
\end{equation}
and $ A_{ext} $ is solution to

\begin{equation}
\frac{d^2 A_{ext} }{d x^2 } + \frac{1}{x} \frac{d A_{ext} }{d x} = 0 .
\end{equation}

We assume that, for $ {\rho }_0 \ll \rho \ll L $, the solution of eq.(22) must
coincide with the expression for the potential generated by a current $ i_T $
flowing through a one-dimensional wire 

\begin{eqnarray}
&& A_w ( \rho ) \simeq 2 \, \frac{i_T }{4 \pi {\epsilon }_0 c^2 } {\int }_0^{L/2}
dz ( {\rho }^2 + z^2 )^{-1/2}   \\ \nonumber
&& = \frac{i_T}{2 \pi {\epsilon }_0 c^2 } \: \ln
\vert \frac{L}{2 \rho } + ( 1 + \frac{L^2 }{4 {\rho }^2 } )^{1/2} \vert \simeq
\frac{i_T}{2 \pi {\epsilon }_0 c^2 } \: \ln \frac{L}{\rho} .
\end{eqnarray}
Taking into account the regularity of $ A_c $ for $ x = 0 $, as well as the
continuity of both $ A_c $ and its gradient across the surface $ x = x_0 $, we
obtain

\begin{equation}
A_{int} = - [ i_T / 2 \pi {\epsilon }_0 c^2 x_0 I_1 (x_0 ) ] \, I_0 (x)  ,
\end{equation}
where the $ I $'s are modified Bessel functions (imaginary-argument Bessel
functions) \cite{9}. We obtain also

\begin{equation}
A_{ext} = - \frac{ i_T }{ 2 \pi {\epsilon }_0 c^2 } [ \ln \frac{x}{x_0 } +
\frac{ I_0 (x_0 ) }{ x_0 I_1 ( x_0) } ]  .
\end{equation}
By a comparison of eq.(20), for $ x > x_0 $, with eq.(23) we obtain 

\begin{equation}
i_T = e_F (x_0 ) i_p   ,
\end{equation}
with $ i_p = \pi {\rho }_0^2 \langle j_p \rangle $ and $ e_F $ given by

\begin{equation}
e_F = 2 [ x_0^2 \ln ( X_L /x_0 ) + x_0 I_0 ( x_0 ) / I_1 ( x_0 ) ]^{-1},
\end{equation}
where we have put $ X_L = \gamma L $. Physically eq.(26) is a consequence of
the skin-effect. A plot for $ e_F $ versus $ x_0 $ is given in Fig.1, for $ X_L
/x_0 = 10^4 $. \par

\begin{figure}
\onefigure{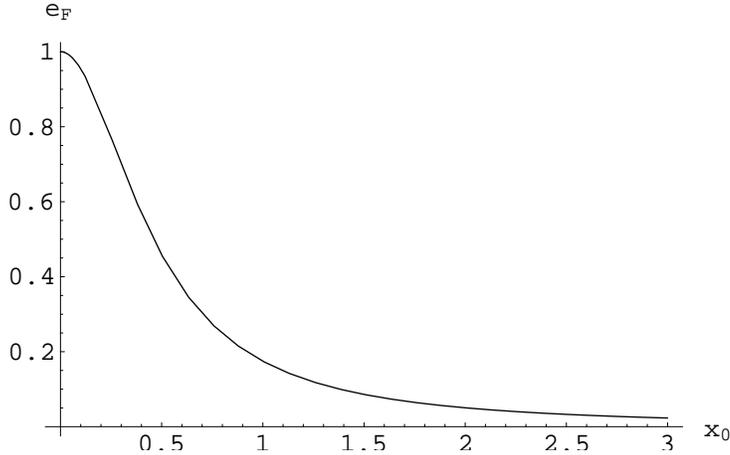}
\caption{Plot of $ e_F $ versus $ x_0 $, for a conducting wire .The function $
e_F $ represents the ratio between the total current $ i_T $ and the canonical
one $ i_p $.The reduced radius $ x_0 = \gamma {\rho }_0 $ represents the 
radius of the wire in units with $ {\gamma }^{-1} = 1 $, where  $ {\gamma
}^{-1} $ is the {\it penetration depth}.  For copper {\it e.g.}  $ x_0 = 1 $
corresponds to a radius $ {\rho }_0 \simeq 1.8 \, \times \, 10^{-8} $ m.
Corrections due to surface effects can be expected. The plot is given for $ L/
{\rho }_0 = 10^4 $, where $ L $ is the length and $ {\rho }_0 $ the radius of
the conducting wire.}
\label{fig1} 
\end{figure}

According to eq.(12) the magnetic energy $ {\cal E}_F $ is given by

\begin{equation}  
{\cal E}_F \simeq - ( N {\hbar }^2 p^2 / 2 m) [ 1 - e_F (x_0 ) ]
 \simeq  - (N {\hbar }^2 p^2 / 2 m )  .
\end{equation}
The last expression in eq.(28) represents the asymptotic limit for $ x_0 \gg 1
$ ({\it bulk limit}). It is independent of geometric details and coincides with
the r.h.s. of eq.(17) obtained for the previous model. Such a coincidence
ascribes a sort of universality to this result. In the derivation of eq.(28)
the following identity

\begin{equation}
{\int }_0^{x_0 } dx \: x I_0 ( x ) = x_0 I_1 ( x_0 ) 
\end{equation}
has been used.  \par

\section{Conclusions}

Let $ {\cal E}_M $ be the ground-state energy for a system of electrons, in the
absence of current. Let $ \Delta {\cal E}_M $ be the variation for $ {\cal E}_M
$, corresponding to a translation in momentum space, assuming that such a
translation is allowed by the electron state. According to eq.(12), the
corresponding variation for the total energy of the system is given by

\begin{equation}
\Delta {\cal E } =  {\cal E }_F + \Delta {\cal E }_M  .
\end{equation}
Since the magnetic energy $ {\cal E}_F $ is negative, a physical situation, in
which $ \Delta {\cal E} $ is negative, is possible. In such a case the true
ground state  of the system would be characterized by a spontaneous current, as
supposed in Ref.s \cite{4} and \cite{5}. \par

It can be demonstrated that, as a consequence of eq.(28),  the ground state of
a system of electrons in a periodic potential, with periodic boundary
conditions, is {\it at least} degenerate, up to finite-size effects. In fact,
let $ \vert {\psi }_0 \rangle $ be the ground state of $ H_M $ and let  $ {\cal
E}_M $ be the corresponding eigenvalue. In the state  $ \vert {\psi }_0 \rangle
$ the total canonical momentum, as well as the total current, vanish. Let $
\vert {\psi }' \rangle $ be the state obtained from  $ \vert {\psi }_0 \rangle
$ through a rigid translation in momentum space

\begin{equation}
\vert {\psi }' \rangle = \exp [i {\hbar }^{-1} \bm{p} \cdot {\sum }_j
{\bm{r} }_j ] \; \vert {\psi }_0 \rangle  .
\end{equation}
We obtain

\begin{equation}
\langle {\psi }' \vert  H_M  \vert {\psi }' \rangle = {\cal E}_M + N ( {\hbar
}^2 p^2 / 2m ) ,
\end{equation}
and, according to eq.(28), $ \Delta {\cal E } \simeq 0 $. However this result,
since it holds up to finite-size effects only, is not sufficient to allow a
spontaneous current.  \par

The possibility for $ \Delta {\cal E } < 0 $ is investigated, for the sake of
simplicity, in the case of a system of Bloch electrons in a conducting band.
However a similar analysis could be applied, in principle, to more complicated
electron states. In this simple case we have 

\begin{equation}
{\cal E}_M = ( V / 8 {\pi }^3 )\int d \bm{k} \, n ( \bm{k} ) \varepsilon
 ( \bm{k} )
\end{equation}
and \cite{8}

\begin{equation}
\bm{p} \,  =  ( V / 8 {\pi }^3 N \hbar ) \int d \bm{k} \, n ( \bm{k} ) \langle
{\psi }_{\bm{k} } \vert ( - i \hbar \bm{\nabla} ) \vert  {\psi }_{\bm{k} }
\rangle  = ( m V / 8 {\pi }^3 N {\hbar }^2 ) \int d \bm{k} \, n ( \bm{k} )
{\bm{\nabla}}_{\bm{k} } \varepsilon ( \bm{k} ).
\end{equation}
Neglecting the magnetic energy of eq.(28), in the ground state one would
obtain, for  $ n ( \bm{k} ) $, the zero-temperature Fermi-Dirac  distribution
function

\begin{equation}
n_F ( \bm{k} ) \equiv  n_F ( \varepsilon (\bm{k} ))= 2 \theta (
{\varepsilon }_F - \varepsilon ( \bm{k} ) ) 
\end{equation}
and correspondingly $ \bm{p} = \bm{0 } $ and vanishing current. However, owing
to the negative value of $ {\cal E}_F $, it may happen that the solution given
in eq.(35) is unstable, when the magnetic interaction is switched on. In fact
let us consider the following distribution function

\begin{equation}
n( \bm{k} ) = n_F ( \bm{k} - {\bm{e} }_z \, q) ,
\end{equation}
obtained from $ n_F $ through a shift of the argument. We observe that the
transformation of the distribution function given in eq.(36) represents a sort
of {\it energy-weighted} translation in momentum space, not a rigid one as
given in eq.(31). In this case we have

\begin{equation}
p \,  = - \frac{m V q}{8 {\pi }^3 N {\hbar }^2 }  \int d \bm{k} \, \frac{d n_F
}{d \varepsilon }( \frac{\partial \varepsilon }{\partial k_z } )^2   =  
\frac{m V q}{4 {\pi }^3 N {\hbar }^2 }  \int d \bm{k} \, \delta  ( \varepsilon 
- {\varepsilon }_F ) ( \frac{\partial \varepsilon }{\partial k_z } )^2
\end{equation}
and

\begin{equation}
\Delta {\cal E}_M  = - \frac{ V q^2 }{16 {\pi }^3 }  \int d \bm{k} \, \frac{d
n_F }{d \varepsilon }( \frac{\partial \varepsilon }{\partial k_z } )^2  =
\frac{N {\hbar }^2 }{2 m } \, p q ,
\end{equation}
where the parity of $ \varepsilon ( \bm{k} ) $ has been used. Now eq.(30) 
reads

\begin{equation}
\Delta {\cal E}  \simeq ( N {\hbar }^2 / 2 m ) \, p ( q - p )
\end{equation}
and the condition for $ \Delta {\cal E} < 0 $  consists of  $ p > q $, {\it
i.e.} 

\begin{equation}
(V / 4 {\pi }^3 N) \int d \bm{k} \, \delta (\varepsilon  - {\varepsilon }_F )
\, ( \partial \varepsilon  / \partial k_z  )^2 > ( {\hbar }^2 / m ),
\end{equation}
where $ m $ is the physical mass of the electron. \par 

In the simple case of a spherically symmetric $ \varepsilon $ we obtain from
eq.(40) 

\begin{equation}
k_F^{-1} \vert d \varepsilon / d k {\vert }_{ k = k_F } > ( {\hbar }^2 / m ),
\end{equation}
where the expression $ k_F^3 = 3 {\pi }^2 N V^{-1} $ has been used for the
Fermi momentum $ k_F $. This is the condition for a superconductivity {\it
\`{a} la} Heisenberg \cite{4}. Finite-size effects could destroy the
superconductivity {\it e.g.} in the case of a very small ring, according to
eq.(17), or of a very thin wire, according to eq.(28). \par

For a free-electron gas in the bulk limit, the l.h.s. and the r.h.s. of eq.(41)
coincide. This result indicates that, for such a system, the ground state is
degenerate, up to finite-size effects, provided that the magnetic
self-interaction energy is taken into account properly.


\begin{thebibliography}{0}

\bibitem{1}
  \Name{Bloch F. \and Nordsieck A. }
  \REVIEW{Phys. Rev. }{52}{1937}{54}.

\bibitem{2} 
  \Name{Chung V. }
  \REVIEW{Phys. Rev. B }{140}{1965}{1110}.

\bibitem{3}
  \Name{Jackson J. D. }
  \Book{Classical Electrodynamics, 3rd ed.}
  \Publ{John Wiley, New York}
  \Year{1999}.
  
\bibitem{4}
  \Name{Heisenberg W. }
  \REVIEW{Z. Naturforsch. a }{2}{1947}{185} ; \SAME{3}{1948}{65}.
 
\bibitem{5}
  \Name{Born M. \and Cheng K. C. }
  \REVIEW{Nature}{161}{1948}{968} ; \SAME{161}{1948}{1017}.
 
\bibitem{6}
  \Name{Bohm D. }
  \REVIEW{Phys. Rev.}{75}{1949}{502}.
 
\bibitem{7}
  \Name{Miglietta F. }{ arXiv:}{quant-ph}{/0202049}. 

\bibitem{8}
  \Name{Ashcroft N. W. \and  Mermin N. D. }
  \Book{Solid State Physics}
  \Publ{Saunders College, Philadelphia}
  \Year{1976}.
 
\bibitem{9}
  \Name{Abramowitz M. \and Stegun I. A. }
  \Book{Handbook of Mathematical Functions}
  \Publ{Dover, New York}
  \Year{1965}.



  


\end{thebibliography}
\end{document}